# DBSR: Dynamic base station Repositioning using Genetic algorithm in wireless sensor network

Amir Mollanejad[1], Leili Mohammad Khanli[2] and Mohammad Zeynali[3]

[1] Islamic Azad University- Jolfa
Branch, Iran

[2] Department of computer science  University of Tabriz, Iran

[3] Islamic Azad University-Bostanabad
Branch, Iran

**Abstract**
Wireless sensor networks (WSNs) are commonly used in various ubiquitous and pervasive applications. Due to limited power resources, the optimal dynamic base station (BS) replacement could be Prolong the sensor network lifetime.
In this paper we'll present a dynamic optimum method for base station replacement so that can save energy in sensors and increases network lifetime. Because positioning problem is a NP-hard problem [1], therefore we'll use genetic algorithm to solve positioning problem. We've considered energy and distance parameters for finding BS optimized position. In our represented algorithm base station position is fixed just during each round and its positioning is done at the start of next round then it'll be placed in optimized position. Evaluating our proposed algorithm, we'll execute DBSR algorithm on LEACH & HEED Protocols.

## 1. Introduction

Networking unattended wireless sensors is expected to have significant impact on the efficiency of many civil and military applications, such as disaster management, environment monitoring, combat field surveillance and security [2][3][4]. A wireless sensor network consists of tiny sensing devices, which normally run on battery power. Sensor nodes are densely deployed in the region of interest. Each device has sensing and wireless communication capabilities, which enable it to sense and gather information from the environment and then send the data and messages to other nodes in the sensor network or to the remote base station. Considering the limited energy capabilities of an individual sensor, a sensor node can sense only up to very limited area, so a wireless sensor network has a large number of sensor nodes deployed in very high density (up to 20nodes/m), Which causes severe problems such as scalability, redundancy, and radio channel contention[6].

In this paper we'll find optimized position of Base Station toward the available node in network, we will try the node can gather data and send it to BS with the least possible energy usage. Finding BS optimized position; we've considered energy and distance parameters. BS optimized positioning is a NP-hard problem. Therefore we'll use genetic algorithm to solve positioning problem
The rest of this paper is organized as follows: In the next section we will point out related work; Section 3 describes the network model and assumptions, in section 5 we will discuss proposed algorithm; Section 6 presents simulation results and performance evaluation the conclusion and future work's presented in sections 7.

## 2. Related Work

Attempts to reduce energy usage in wireless sensor networks are one of the most important subjects.
Energy economizing is done by two ways:
1) Using sensors with less energy usage
2) Using power management methods in the design of network software.
For example sending TDMA is suitable in the view of energy usage. Because the sensor is in waiting mode when sensor doesn't send data While Sensors are in this state use the least energy. Also network geometrical configuration methods can reduce energy usage.

Another less considered method is mobility of BS and placing it in a position which is suitable in distance and residual energy better. In this article we will focus on this issue. As we know all of attempts for reducing energy usage in sensors are in order to increase network lifetime.

[5] Presents BS optimized positioning by linear programming. This paper proposes set of procedures to design $(1 − \varepsilon)$ approximation algorithms for base station placement problems under any desired small error bound $\varepsilon > 0$. It offers a general framework to transform infinite search space to a finite-element search space with performance guarantee. BS is not dynamic in this method; we'll suppose that the BS is dynamic.

[1] There is an algorithm presented based on traffic density factor. Their approach tracks the distance from the closest hops to the base-station and the traffic density through these hops. When a hop that forward high traffic is exceeded threshold the base-station qualifies the impact of the relocation on the network performance and moves if the overhead is justified.

## 3. Network Model and Assumption

It's supposed that the network environment is in a two Dimensions space with a specific width and length and sensors are placed in positions with a specific width and length. BS is able to depart and changes its position. It's supposed that at the end of each round, sensors can declare their residual energy to BS. BS departs based on sensors residual energy and distance, so it is placed on an optimal position. It means that it's placed in the nearest position toward all of sensors. Considering, sensors residual energy parameter is effective on the position of BS. BS will be near to sensors whit less residual energy. For example, as shown in figure 1, let's consider a sensor network composed of 20 nodes; at the start of first round BS is in optimum position and in the next round BS is near the sensors with lower energy remaining, see figure 2.

Comparing figures we realize that, in second round sensors lose less energy transferring data to BS than first round, thus increases the network lifetime. DBSR network model and assumptions is:

- our network is in a two Dimensions space with a specific width and length (200m , 200m)

- The position of sensors is random and they equipped with a GPS set.

- BS is able to move in all of network.

- We have one BS for over the network.

- The residual energy of sensors can be calculated.

- Sensors send their residual energy to BS in each round.

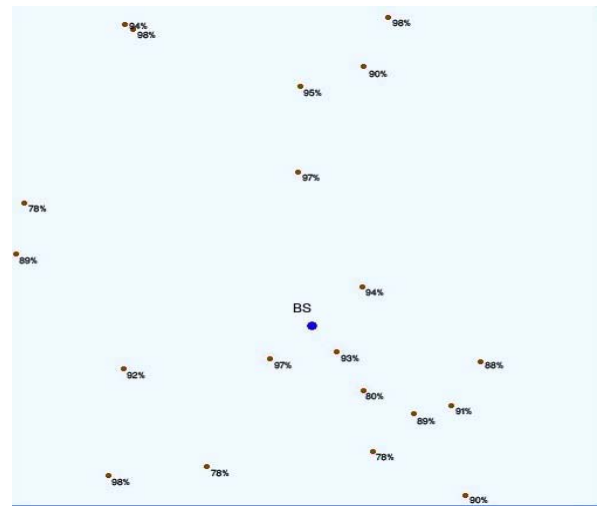

Fig. 1 Base station position in first round

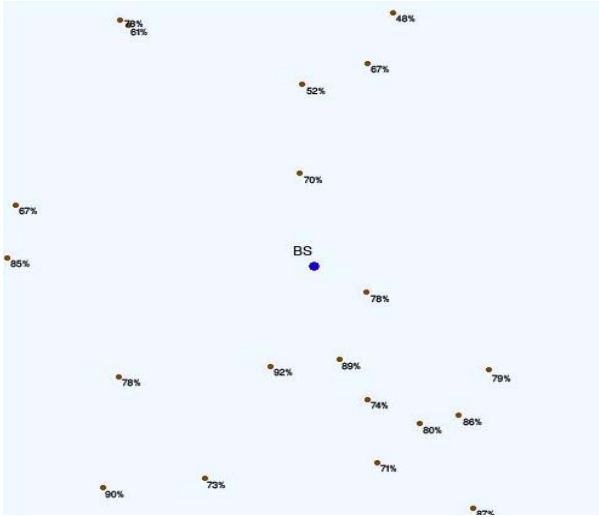

Fig. 2 Base station position in second round

## 4. Genetic Algorithm

Genetic algorithms (GA) are one of the efficient tools that are employed in solving optimization problems [6]. The basic idea of genetic algorithm is as follow [7][8]: the genetic pool of a given population potentially contains the solution, or a better solution, to a given optimization problem. This solution is not active because the genetic combination on which it relies is split between several subjects. Only the association of different genomes can lead to the solution. Optimization in genetic algorithm is based on optimization of a fitness function which is a function of environment individuals or genes. Each new generation is generated by applying Crossover and Mutation operand on old generation. Then in new generation good genes that lead to better fitness function have more chance to survive. So, after some generations the optimal solution will be attained.

## 5. Proposed Algorithm

In DBSR algorithm the primal population consists of n chromosomes which show the position of BS. Each chromosome includes two parts; X (length of network environment) & Y (width of network environment). They have encoded by binary encoding scheme. Each chromosome is evaluated by fitness function. We have applied modified 2-point crossover and random point flip for mutation operation. In additional, for new population replacement, we will replace selected population with next population. The condition of genetic algorithm expiry is based on the number of generations we've supposed. Proposed algorithm pseudo-code is shown below:

For each round
BS receive residual energy massage from all nodes then
BEGIN GA
gen:=0 { generation counter }
Initialize population P(gen)
Evaluate population P(gen)
For gen=0 to n do
gen:=gen+1
Select P (gen) from P (gen -1)
Crossover P (gen)
Mutate P(gen)
Evaluate P(gen)
END FOR
Output best answer
END GA

### 5.1 Population

We have applied binary encoding in our proposed algorithm that is each chromosome is related to BS position. We suppose length and width for the environment which sensors are distributed. It has supposed that all of the sensors are placed in a point with a specific width and length. Chromosomes are consisted of two parts: First binary part is related to X (length of sensor point) and the second binary part is related to Y (width of sensor point).
The number of X & Y bits depends on the length and the width of network environment. If (length=width=200) then to show each one of X & Y we need 8 bits. for instance, randomly generated chromosome represent point X=170, Y=109 see (figure3):

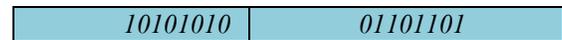

| *10101010* | *01101101* |

Fig 3.  X=170, Y=109

### 5.2 Fitness

Fitness function is calculated based on distance and residual energy parameters in sensors. Each chromosome which enjoys random X & Y that it shows the position of BS. Summation of distance between this random point and all of the sensors is achieved by multiply a ratio for each sensor (this ratio introduces inverse of residual energy in sensor)
that shown in equation (1).

Residual energy is supposed as a number between 1 and 10.

$$obf = \sum_{i=1}^{n} \frac{1}{W_i} \sqrt{(x - x_i)^2 - (y - y_i)^2} \qquad (1)$$

Where $n$ = number of sensor nodes
The fitness function is given as follows:

$$fitness = m - obf \qquad (2)$$

Table 1: The parameters used in equation 1 and fitness function

| Parameter | Description |
|---|---|
| $i$ | Index of nodes |
| $x_i$ | position length of node of i |
| $y_i$ | position width of node of i |
| $W_i$ | Residual energy node of i |
| $m$ | A very large number |

5.3 Selection

The selection process selects chromosomes from the mating pool according to the survival of the fittest concept of natural genetic system. In each successive generation, a proportion of the existing population is selected to breed a new generation. Our approach uses 80% as crossover probability, which means that 80% of the population will take part in crossover. The probabilities for each chromosome are calculated according to their fitness values, and selection is in proportion to these probabilities where the chromosome with lower probability has more chance of being selected. The proportions are calculated as given below.

$$prob(ch_i) = \frac{fitness(ch_i)}{\sum_{i=1}^{n} fitness(ch_i)} \qquad (3)$$

Once the probabilities are calculated, Roulette Wheel selection [9] is used to select parents for crossover.

5.4 Crossover and Mutation

We've used modified two point's crossover for crossover operation that selects two cut points for each of two chromosomes. One cut point for Y part and a cut point for X part are supposed. These points are selected randomly. Crossover operation is done as shown in figure 4.

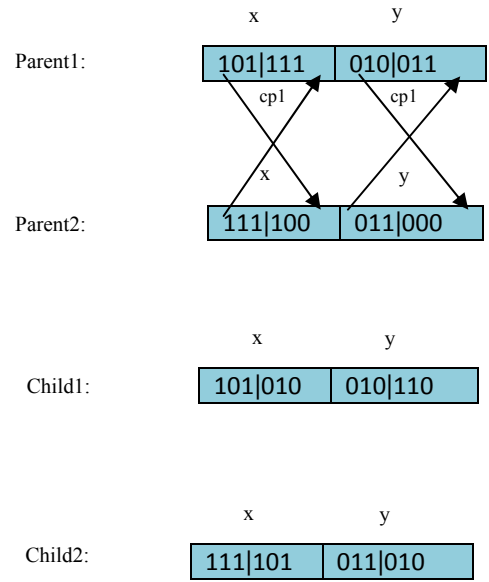

Fig4. Crossover Example

For mutation we select two random points on chromosome. One part X and the other part for Y. We flip the randomly selected bits.

## 6. Simulation and Result

In this part the performance of presented algorithm on LEACH and HEED protocols is evaluated. At first we have executed HEED and LEACH protocols without using the suggested algorithm. At the second stage we have executed the DBSR on HEED and LEACH protocols then we compare the results.

6.1 Sensor network simulation parameter

For these experiments, a network of $N$ sensor nodes in an $200 \times 200 \, m^2$ area is considered. The N nodes are assumed to be uniformly distributed over the area, every simulation result shown below is the average of 100 independent run where each run uses a different randomly-generated population. All parameters are given in Table 2. A simple radio model that also can be found in [10] has been adopted.

Table 2: Simulation Parameters

| Parameters | Description | Value |
|---|---|---|
| $M*M$ | Simulation Area | (0,0)~(200,200) |
| N | Number of | 200 |

| | Node | |
|---|---|---|
| $P_s$ | Sink position | Dynamic for each round |
| rs | Sensing radius rs | 15 m |
| $E_i$ | Initial energy | 1J |
| $L_{data}$ | Data packet size | 200 Bytes |
| Eelec | electronics energy | 50 nJ/bit |
| efs | free space coefficient | 10 nJ/bit/m2 |

### 6.2 GA simulation parameter

The simulation parameters for GA are as follows: a) population size and the number of generations are equal to the number nodes, b) mutation rate is 0.09, c) crossover rate is 0.80, and d) Roulette Wheel selection probability is 0.90.

### 6.3 Results

Figure. 5 shows the total residual energy of the network in two protocols for 20 rounds, with the number of node 200. It shows that HEED with DBRS balances the energy consumption among all nodes best.

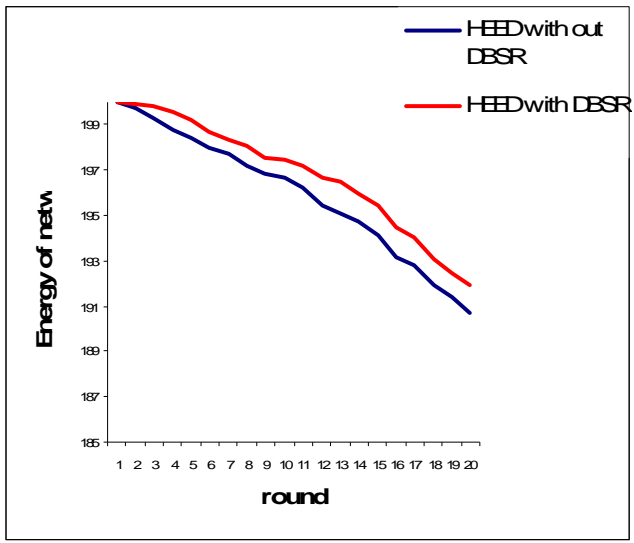

Fig 5. The total residual energy of the network

Figure. 6 shows the total residual energy of the 200 node network in two protocols for 20 rounds, It shows that LEACH with DBRS balances the energy consumption among all nodes best.

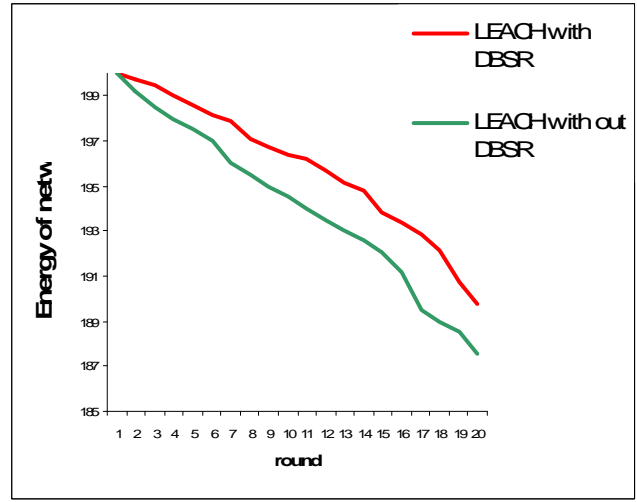

Fig 6. The total residual energy of the network

Figure. 7 illustrate simulation results of our sample network. We compare the original LEACH algorithm with LEACH - DBSR. For *First Node Dies* (FND) [16] a 35% improvement is accomplished comparing the LEACH - DBSR algorithm with original LEACH. *Half of the Nodes live* (HNA) [16] improves by 36 %.

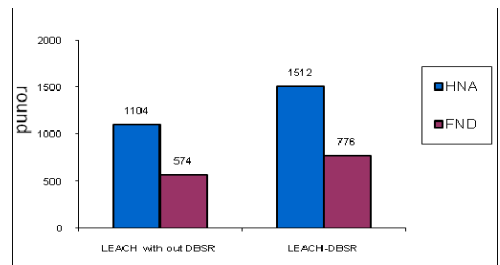

Fig 7. network life time comparison using FND and HNA criteria's between LEACH, LEACH-DBSR.

Figure. 8 illustrate simulation results of our sample network. We compare the original HEED algorithm with HEED-DBSR. For FND a 38% improvement is accomplished comparing the HEED with DBSR algorithm with original HEED. HNA improves by 22 %.

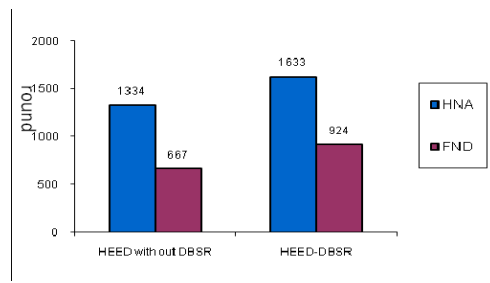

Fig 8. network life time comparison using FND and HNA criteria's between HEED, HEED -DBSR.

## 7. Conclusion

In this paper we introduce dynamic optimized positioning method for BS optimized positioning. That can save energy in sensors and increases network lifetime. We applied genetic algorithm, for dynamic optimum BS replacement. Simulation results show that, DBSR outperforms other schema significantly in optimizing sensor's energy consumption and improving network lifetime. In future work we can use learning automata for dynamic optimum BS replacement.

## References

[1]. M. Younis, A. Lalani, M. Eltoweissy, "Safe base-station repositioning in wireless sensor networks," pcc, pp.70, 2006 IEEE International Performance Computing and Communications Conference, 2006

[2]. I. F. Akyildiz et al., "Wireless sensor networks: a survey", *Computer Networks*, Vol. 38, pp. 393-422, 2002.

[3]. D. Estrin, et al., "Next Century Challenges: Scalable Coordination in Sensor Networks," in the *Proceedings of*

4]. J.M. Rabaey, et al., "PicoRadio supports ad hoc ultra low power wireless networking," *IEEE Computer*, Vol. 33, pp. 42-48, July 2000.

[5] Yi Shi**,** Y. Thomas Hou, and Alon Efrat, "Algorithm design for a class of base station location problems in Sensor Networks," ACM/Springer Wireless Networks, vol. 15, issue 1, pp. 21-38, 2009.

[6]. B. Thomas, F. Hoffmeister, "*Global optimization by mans of evolutionary alghorithms*" in random        Search as Method for Adaptation and Optimization of Complex Systems, edited by: A. N. namoshkin, Kras-Nojarsk Space Technology University, pp. 17-21, 1996

[7]. Goldberg D., *Genetic Algorithms,* Addison Wesley, 1988.

[8]. Holland J.H., *Adaptation in natural and artificialsystem,* Ann Arbor, The University of Michigan    Press, 1975.

[9].    Fitness    Proportionate    Selection    (2007), http://en.wikipedia.org/wiki/Fitness_proportionate_selection

[10] I. Texas Instruments, "MSP430x13x, MSP430x14x Mixedm Signal Microcontroller. Datasheet, " 2001.